\documentclass{iaus}
\usepackage{graphicx}

\title[Cloudy_3D] 
{Cloudy\_3D, a new pseudo-3D photoionization code}

\author[C. Morisset]   
{Christophe Morisset$^1$ }

\affiliation{$^1$Instituto de Astronomia, UNAM, Mexico\break email:Morisset@astroscu.UNAM.mx}

\pubyear{2006}
\volume{234}  
\pagerange{xxx--xxx}
\date{?? and in revised form ??}
\jname{Proceedings Title IAU Symposium}
\editors{A.C. Editor, B.D. Editor \& C.E. Editor, eds.}

\begin{document}

\maketitle

\begin{abstract}
We developed a new quick pseudo-3D photoionization code based on Cloudy \cite{GF} and IDL (RSI) tools.
The code is running the 1D photoionization code Cloudy  various times, changing at each run the input parameters (e.g. inner radius, density law) according to
an angular law describing the morphology of the object.
Then a cube is generated by interpolating the outputs of Cloudy. In each cell of the cube, the physical conditions (electron temperature and density, ionic fractions) and
the emissivities of lines are determined. Associated tools (VISNEB and VELNEB\_3D) are used to rotate the nebula and to compute surface brightness maps and emission
line profiles, given a velocity law and taking into account the effect of the thermal broadening and eventually the turbulence. Integrated emission line profiles are computed,
given aperture shapes and positions (seeing and instrumental width effects are included).
The main advantage of this tool is the short time needed to compute a model (a few tens minutes).
\end{abstract}

\firstsection 
\section{Introduction}
Most planetary nebulae are not spherically symmetric. Therefore 3D photoionization codes are very welcome. The problem with truly 3D codes (such as Mocassin, \cite{BE}) is their large computing time, which makes it difficult to explore a large parameter space. Morisset et al. (2005) have developed a pseudo-3D code based on the 1D photoionization code NEBU \cite{P02}. Some applications are discussed in \cite{M05} and \cite{M06}. The advantages and limitations of pseudo-3D codes are discussed in those papers as well.

Using the same approach as for NEBU\_3D, we developed Cloudy-3D, a pseudo-3D code based on the widely used 1D code Cloudy \cite{GF}. Here we describe the use of Cloudy\_3D and give information on how to install the software.

\section{Flow diagram of Cloudy\_3D}
In the following we describe the various steps needed to compute a 3D model. 
\begin{itemize}
\item {\bf Definition of the morphology}
Define all the parameters that change with the
angle using angular law, (e.g.
the inner radius is defined as an ellipsoidal,
the hydrogen density is proportional to 1/r$^2$).
\item {\bf Prepare\_Cloudy\_in}
Generate of the N input files for Cloudy, according
to the morphology and the definition
of fixed parameters (e.g. the ionizing flux description,
kept constant for each direction)

- Inputs: Description of the constant and varying parameters
(angular law)

- Outputs: A set of N Cloudy input files
\item {\bf Cloudy\_driver}
Run N times Cloudy, manage the output naming.

- Inputs: A set of N Cloudy input files

- Outputs: A set of N Cloudy outputs files
(physical parameters, line emissivities, etc).
\item {\bf Cloudy\_3D}
Read the N outputs of
Cloudy and interpolate into a
3D coordinates cube.

- Inputs: A set of N Cloudy outputs files; the geometrical size of the coordinate
cube (can be 2D for testing, can be a close-up on one part of the nebula)

- Outputs: The 3D cube containing the Ne, Te, the emissivities of selected lines and the ionic fractional abundances in each cell.
\item {\bf VELNEB\_3D (optional)}
Compute the line profiles given a velocity field.

- Inputs: The description of the velocity field; the cube of emissivities and Te (could come from another 3D code)

- Outputs: The emission line profile in each cell of the 3D cube.
\item {\bf VISNEB\_3D}
Turn the nebula, generate the outputs like the emission maps (3 colors), PV-diagrams, etc.

- Inputs: The cube of emissivities and physical parameters (could come
from another 3D code); rotations to apply to the 3D data; optionally: cube of emission line profiles; position and size of apertures, seeing, instrumental resolution

- Outputs: Surface brightness maps; diagnostic line ratios maps; emission line profiles through apertures, PV-diagrams.
\end{itemize}

\section{What do you need to run Cloudy\_3D?}

\begin{itemize}
\item Cloudy (!), from G. Ferland
\item IDL (RSI)
\item Subversion (svn), used to download the package and to keep it synchronized with the ultimate repository version.
\item E\_mail me (morisset@astroscu.unam.mx) to ask for an account to download the package.
The website dedicated to Cloudy\_3D is in its very beta version, you can register at http://132.248.1.102/Cloudy\_3D
\end{itemize}

\section{Further developments}
\begin{itemize}
\item The catalogue of emission line profile (see Morisset \& Stasinska, this conference).

\item The present version of Cloudy\_3D is asymmetrical (no longitude dependence), the next step of the development if to have a full 3D code.

\item Interfacing Cloudy\_3D with SHAPE (W. Steffen, this conference) will give the possibility to explore quickly the morphological parameter space.

\item Implementation of Cloudy\_3D into a virtual observatory to allow users to run 3D models from the web and to access a database of the previously run models.
\end{itemize}

\begin{acknowledgments}
The computations are made on a AMD-64bit computer financed by grant PAPIIT IX125304 from DGAPA (UNAM, Mexico).

C.M. is partly supported by grant Conacyt-40095 (Mexico).
\end{acknowledgments}

\end{document}